# Isolating Polaritonic 2D-IR Transmission Spectra

Rong Duan[(1)], Joseph N. Mastron[(1,2)], Yin Song[(2,#)], Kevin J. Kubarych[(1)]*
[1] Department of Chemistry, University of Michigan, 930 N. University Ave, Ann Arbor, MI 48109
[2] Department of Physics, University of Michigan, 430 Church Ave, Ann Arbor, MI 48109
[#] Present address: Beijing Institute of Technology, 5 South Zhong Guan Cun St., Beijing, China

*Email: kubarych@umich.edu

**Abstract:** Strong coupling between vibrational transitions in molecules within a resonant optical microcavity leads to the formation of collective, delocalized vibrational polaritons. There are many potential applications of "polaritonic chemistry," ranging from modified chemical reactivity to quantum information processing. One challenge in obtaining the polaritonic response is to remove a background contribution due to the uncoupled molecules that generate an ordinary 2D-IR spectrum whose amplitude is filtered by the polariton transmission spectrum. We show that most features in 2D-IR spectra of vibrational polaritons can be explained by a linear superposition of this background signal and the true polariton response. Through a straightforward correction procedure, where the filtered bare molecule 2D-IR spectrum is subtracted from the measured cavity response, we recover the polaritonic spectrum.

Strong coupling between electronic states of matter and resonant optical cavities has played a key role in the development of novel states of quasi-particles, such as Bose-Einstein condensation[1] and lasing in exciton-polaritons.[2-3] Very recently there has been a sharp growth in chemical applications of strong coupling, led by the discoveries of Ebbesen *et al.* that chemical reactivity can be altered by vibrational strong coupling.[4-9] Infrared spectroscopy of vibrational strong coupling offers direct access to the polaritonic states, and ultrafast pump-probe or 2D spectroscopy enables determination of relaxation and dephasing in these novel hybrid light-matter states.[10-20] The promise of cavity controlled chemistry is sufficiently great that intense investigation is clearly warranted.[21-32]

When $N$ molecules are resonantly coupled to a cavity mode, the eigenstates of the composite system consist of two highly delocalized, collective states that carry both photonic and matter character.[33-34] These polaritons are separated in energy by the so-called vacuum Rabi splitting, which is proportional to the scalar product of the transition dipole moment and the cavity wave vector, as well as the square-root of the number of molecules in the cavity. In addition, there are $N-1$ "dark" states, that resemble the uncoupled molecules (also referred to as the "bare molecules" to indicate a cavity-free condition). Because the cavities required to support a mid-IR wavelength are at least $\lambda/2$, $N$ is typically very large (>$10^{10}$), leading to a large number of dark states relative to the two polaritons. However, the polaritons have a very large oscillator strength due to the participation of $N$ molecules and the cavity mode. A quantitative description of the interaction of the polaritons and the dark states is central to understanding vibrational polariton relaxation and dephasing, as well as the mechanisms by which chemical reactions can be controlled.[14, 21, 35-37] In addition to the dark states, there is also a population of "uncoupled" molecules that do not participate in the delocalized polariton due to their spatial position along the cavity at regions of low field mode amplitude (see discussion in the SI). In terms of frequency, both the dark and uncoupled molecules are essentially indistinguishable from ordinary molecules in the absence of a cavity, though it remains to be determined if the dark states are distinct dynamically from the cavity-free bare molecules.[12, 14, 38]

The fundamental spectroscopy and dynamics of vibrational polaritons present new avenues in understating and manipulating quantum dynamics using the external influence of a cavity. Several recent reports of transition metal complexes, particularly the strong triply-degenerate $W(CO)_6$ stretching vibrations, coupled to resonant cavities have considered the nonlinear optical response of vibrational polaritons in various conditions of weak and strong coupling.[10-19, 38-40] A key proposal to explain the observed relaxation dynamics posits that the polariton states relax to the uncoupled reservoir molecules. Subsequent excited state absorption from the $v = 1$ to $v = 2$ state of these reservoir molecules is thought to account for the substantial signal amplitude at the lower polariton detection frequency due to a coincidental energy matching of the 1-2 transition and the LP frequency. Other work has considered that there is a cavity-length tunable nonlinear response that could potentially be used for polaritonic switching devices.[15] A recent report of cavity-coupled sodium nitroprusside, which has a single NO nitrosyl vibrational transition, was investigated in order to eliminate possible complications of involving three triply-degenerate modes of $W(CO)_6$.[38] In general, the spectra appear actually quite similar and indicate that there are unifying aspects of single vibrational mode polaritons.

In the sodium nitroprusside (SNP) example, Grafton *et al*. present a method to subtract the "reservoir" 2D-IR spectrum due to the uncoupled molecules.[38] Using the response at a 25-ps waiting time ($t_2$) delay, which is longer than the cavity lifetime, they generate a reservoir-only spectrum by extracting two slices of the 2D-IR spectrum: one at an excitation frequency matching the bare molecule peak maximum, and the other at a detection frequency corresponding to the maximum in the 2D spectrum. The product of these two slices is a 2D surface that resembles the overall appearance of the 25-ps spectrum. This spectrum is then subtracted from the early waiting time (1 ps) 2D-IR spectrum of the cavity response, yielding a putative reservoir-subtracted 2D-IR spectrum. A similar analysis was also used to correct the pump-probe transient absorption spectrum.

This method of estimating a 2D spectrum is similar to assuming a purely homogeneous limit, though we are not aware of a precedent for constructing a 2D-IR spectrum by multiplying two slices at one waiting time delay. Khalil *et al*. found that SNP exhibits substantial inhomogeneous broadening as well as spectral diffusion on the same time scales as the polariton studies.[41] Subtracting the constructed homogeneous reservoir spectrum from earlier time 2D spectra, where inhomogeneous broadening leads to diagonally correlated line shapes, cannot capture the dynamic peak shape of the uncoupled molecules. Using a model response function, we show that the background-subtracted 2D spectrum reported previously results from this dynamic 2D line shape in the filtered uncoupled response (details are in the Supporting Information, SI). Nevertheless, the insight that it is necessary to subtract a background 2D response due to the uncoupled molecule is essential for obtaining the polariton response. Our approach takes into account the dynamic line shape by recording the 2D-IR spectrum of the cavity-free sample and multiplying the spectrum at each waiting time by the two-dimensional polariton transmission to account for the spectral filtering of both the excitation and the detection processes. We show below that this approach is capable of isolating the polaritonic 2D-IR response.

If one neglects the dark states, it is possible to view the polariton system as one oscillator coupled to one cavity mode.[33-34] It is common to invoke an effective coupling that collects the electric dipole interaction as well as the square root of *N* dependence. A recent treatment by Herrera *et al*. includes explicitly the state dependent transition dipole moments of an anharmonic oscillator, to arrive at a set of energy levels of the polaritonic ladder beyond the first excited state manifold.[42] With the energy level parameters determined using this method and reported in Grafton *et al*.,[38] we use conventional response functions[43] to produce a purely polaritonic 2D spectrum (**Fig. 1**). Keeping in mind that the signal phase in the polariton case is opposite that in the uncoupled background: ground state bleach (GSB) and stimulated emission (SE) pathways in the polariton lead to reduced transmission, whereas these pathways lead to increased transmission in the uncoupled response. The same logic leads to excited state absorption pathways with opposite signs in the polariton and uncoupled response. In our color scheme, the red features are increased transmission (reduced absorption), and the blue features are decreased transmission (increased absorption).

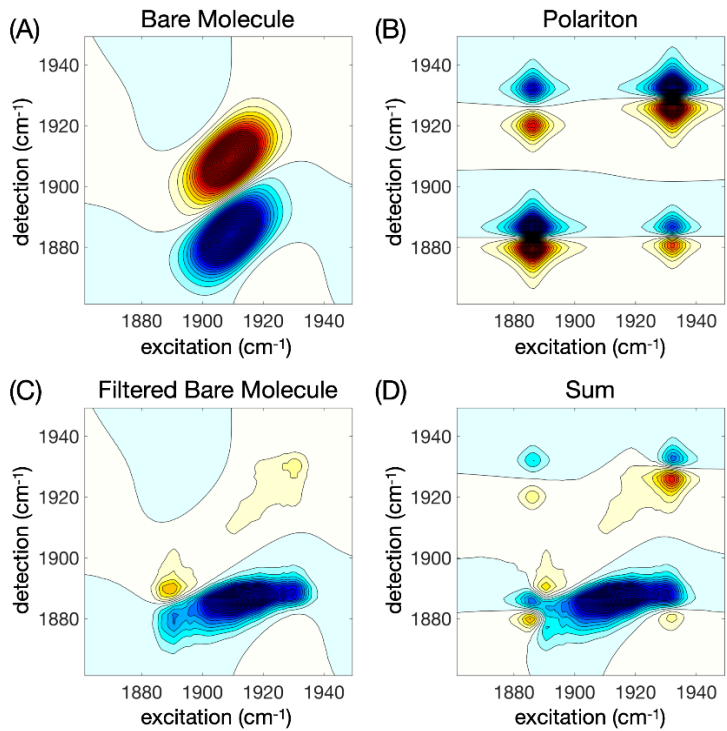

**Figure 1. (A)** Simulated 2D-IR absorptive response characteristic of sodium nitroprusside (SNP) based on the work of Khalil *et al*. **(B)** Polariton 2D-IR spectrum parameterized using the energy levels reported in Grafton *et al*. Note that the features corresponding to ground state bleach and excited state absorption have signs opposite to those of the absorptive bare molecule response. **(C)** Bare molecule response filtered by the polariton transmission (taken from data in Ref. 38). **(D)** Sum of the filtered background and the polariton response.

Using the filtered 2D-IR spectrum (**Fig. 1**), either measured independently, or modeled using a simple response function, we can construct a 2D-IR spectrum for the cavity by adding the uncoupled background to the polaritonic response. We note that we currently do not know *a priori* how to weight the two contributions, but have developed an empirical method to correct our measured spectra (described below). Work by Xiong *et al*. shows that pump-probe spectra change with pathlength at constant concentration,[15] an effect that could be due to changes in the background contribution. The simulations here use weights chosen to resemble the characteristic 2D-IR spectra that are in the literature, as well as the many spectra we have recorded ourselves. The simulations show a striking resemblance to previously reported results. We do note one feature of the spectra that is clearly not reproduced in these simulations is the prominent cross peak at ($\omega_1$ = 1956 cm$^{-1}$, $\omega_3$ = 1977 cm$^{-1}$) corresponding to LP excitation and dark state fundamental detection. This feature could indicate evidence for the LP-to-reservoir energy transfer, and the sign is precisely what would be expected for excited state absorption, but the frequency corresponds to the *v* = 0 to 1 transition of the reservoir. Stimulated emission due to LP-to-dark state energy transfer would have the opposite sign (it would appear red in the present representation). We also find a cross peak at the UP-dark region, but again, it has the incorrect sign to be attributed to stimulated emission.

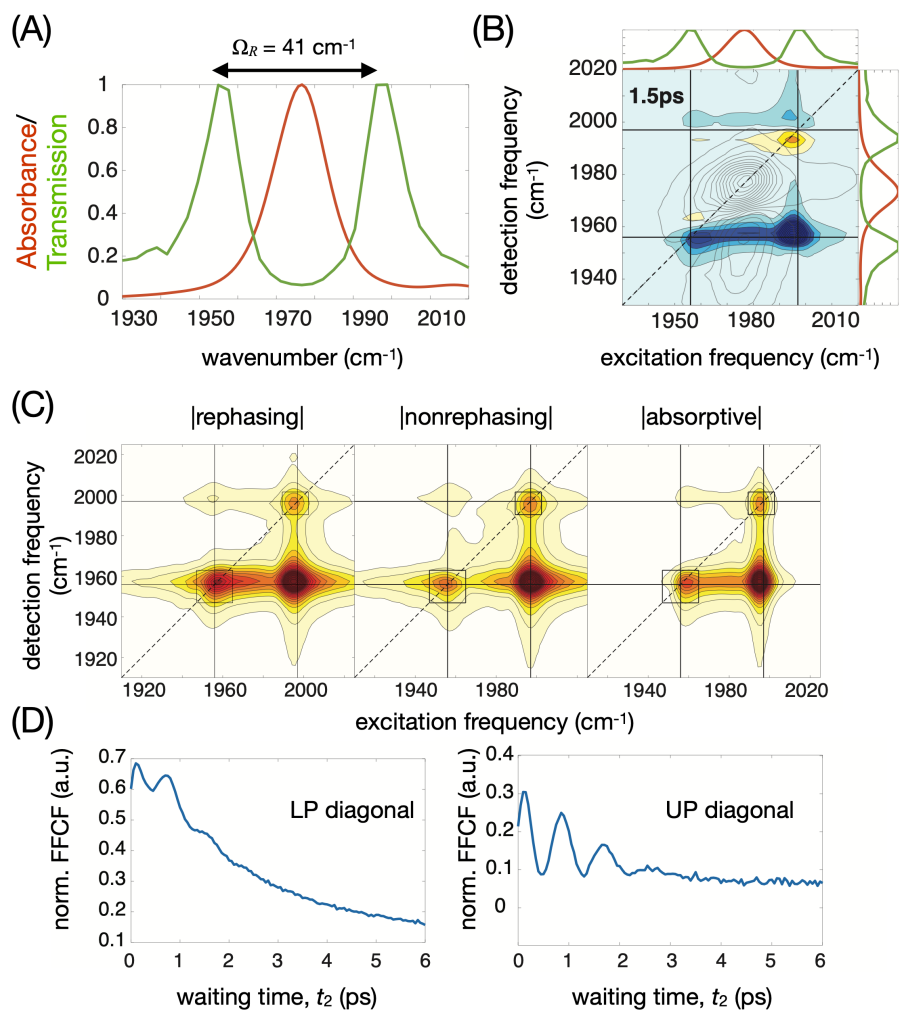

**Figure 2. (A)** Linear absorbance (red) of 2-mM W(CO)$_6$ in butylacetate in a standard cell, and transmission (green) of 40 mM W(CO)$_6$ in butylacetate in a resonant cavity (mirror reflectivity = 92%). Polariton bands are at 1956 and 1997 cm$^{-1}$; the bare molecule band is at 1977 cm$^{-1}$. **(B)** Absorptive 2D-IR spectrum of the cavity response (color) superimposed on the bare molecule response (gray), showing the significant overlap of the v=1 to v=2 excited state absorption and the lower polariton. Linear responses are shown along each axis for reference. **(C)** Absolute value rephasing, nonrephasing and absorptive 2D-IR spectra for the cavity; the signals are obtained using a 6 phase cycling scheme (SI). **(D)** Normalized frequency fluctuation correlation functions (FFCFs) for the LP diagonal (left) and the UP diagonal both indicate coherent modulation due to the nonrephasing contribution. The LP diagonal shows a pronounced spectral inhomogeneity that decays on the same time scale (2.4 ps) as the bare molecule measured outside of a cavity. This apparent spectral diffusion is due to the overlapping uncoupled background 2D-IR response.

Subtracting the uncoupled background 2D-IR requires measuring that response under conditions that are as similar as possible to the cavity-coupled conditions. Due to the typically high concentrations used in establishing strong coupling, there can be distortions to the 2D spectrum arising from solute-solute interactions as well as optical effects such as signal reabsorption. For both the bare molecule and polariton experiments we used ~40 mM W(CO)$_6$ in butyl acetate. To reduce the optical density and signal reabsorption, we used a 6-μm spacer for the bare molecule 2D experiment, whereas the cavity used a 25-μm spacer. In general, an ideal subtraction should be done with identical sample pathlengths and concentrations, but this is often difficult in practice with the cavity lengths required to form the polaritonic system. The FTIR was recorded using a lower concentration sample (2.5 mM) in order to obtain an undistorted linear absorption spectrum. The 2D-IR experimental set up has been described previously, and is summarized in the SI.

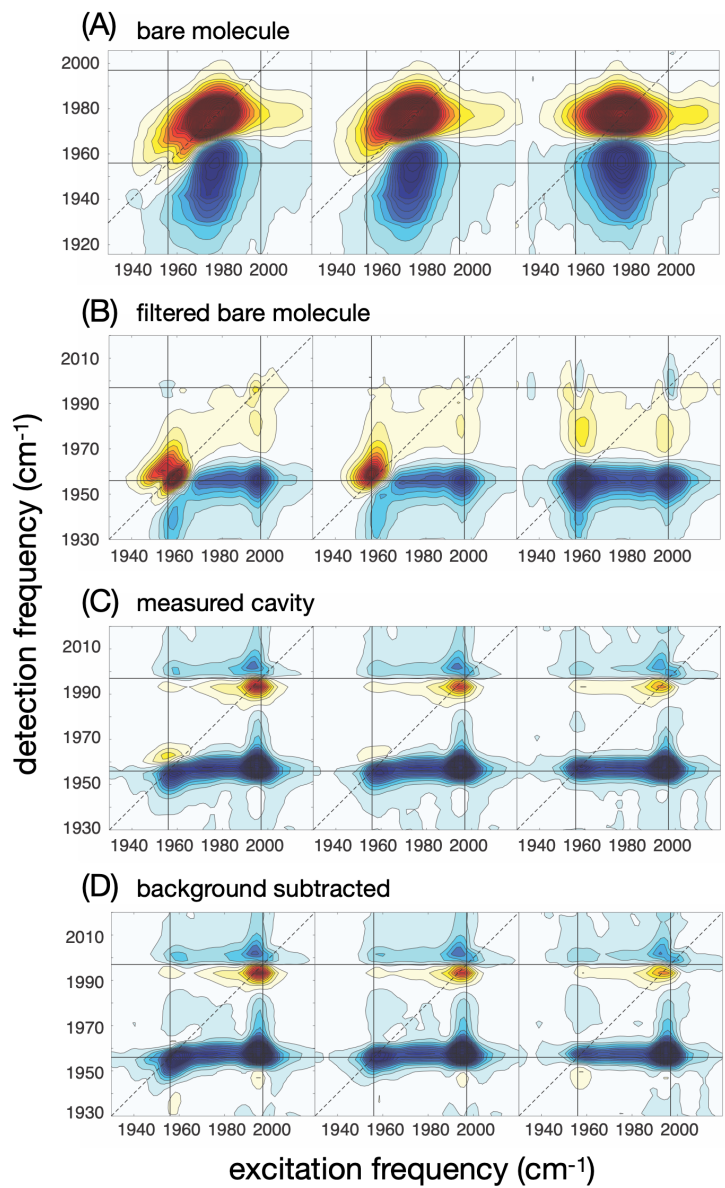

**Figure 3. (A)** Absorptive bare molecule 2D-IR obtained using 40 mM W(CO)$_6$ in butylacetate with a 6 μm pathlength cell (uncoated, 3-mm CaF$_2$ windows). All sets of three 2D spectra are shown for the same three waiting times (0.8, 1.5, and 30 ps). Dynamic changes of the line shape are due to spectral diffusion. **(B)** Absorptive 2D-IR response filtered by the polariton transmission (shown in Fig. 2A). **(C)** Measured absorptive 2D-IR spectra of the full cavity response (40 mM W(CO)$_6$ in butylacetate with a 25 μm pathlength; mirror reflectivity is 92%). **(D)** Background subtracted response (cavity − filtered bare molecule) shows the spectral signatures of the polaritonic ladder expected for an anharmonic oscillator coupled to a single harmonic cavity mode.

The as-measured 2D-IR spectra of the cavity and bare molecule shows significant overlap of the 1-2 bare molecule response and the LP-detected 2D-IR signal (**Fig. 2B**). Absolute value rephasing, non-rephasing and absorptive spectra show prominent peaks at locations expected based on the linear transmission spectrum. Using the inhomogeneity index as a measure of the frequency-fluctuation correlation function,[44] we find coherent oscillations due to the impulsive excitation of both polariton bands, as well as a slow decay in the LP diagonal signal. This decay matches the FFCF decay of the bare molecule (2.4 ps), which arises from spectral diffusion. The polariton bands are purely homogenously broadened, and therefore cannot undergo spectral diffusion. This signature is, therefore, indicative of the uncoupled background 2D-IR signal.

Given the ability to construct 2D spectra of cavity systems by adding a pure polaritonic response to the filtered bare-molecule 2D-IR spectrum, it should be possible to isolate the pure polaritonic response experimentally. In our transmission polariton 2D measurement, the probe, serving as a local oscillator (LO), propagates within the cavity along with the signal, therefore the LO should have the same shape as the polariton transmission spectrum [$T(\omega)$]. The uncoupled bare molecule signal is modulated by the LO [$S_{sig,bare\ molecule} \propto 2\text{Re}(E_{LO}(\omega_{det})E_{sig}(\omega_{det}))$]. The excitation pulses are filtered in a similar way. **Figure 3** shows the process of obtaining polaritonic spectra from the measured cavity response. We record the 2D-IR spectrum of the same sample in an ordinary cell without reflective cavity mirrors. The polariton transmission spectrum, $T(\omega)$, is measured in-situ using the pump and probe pulses (Fig. 2A). By multiplying the bare molecule 2D-IR spectrum by the polariton transmission in both excitation and detection dimensions [i.e. $T(\omega_1)T(\omega_3)$], we obtain the filtered bare molecule response. To obtain the pure polaritonic contribution, we subtract the filtered bare-molecule 2D-IR spectrum from the measured spectrum of the cavity system (i.e. $S_{residual} = S_{cavity} - \alpha S_{filtered\ bare\ molecule}$). We use data with the same waiting time ($t_2$) to account for spectral shape changes that arise from the spectral diffusion within the inhomogeneously broadened band. Given that polaritons are immune to inhomogeneous broadening, we set the weight for subtraction such that we minimize the residual spectral diffusion in the LP diagonal peak (see SI for details). The resulting polaritonic 2D-IR spectra at various waiting times illustrate that the prediction based on two coupled oscillators yields a good correspondence with the measurements.

Several peaks in the polaritonic 2D spectrum exhibit oscillatory behavior with a frequency corresponding to the vacuum Rabi splitting (**Fig. 4**), as is expected in 2D-IR spectra of coupled modes having narrow homogeneous linewidths. This oscillatory behavior is present in other 2D-IR spectra of polaritons,[12, 38] and it is well established that the diagonal peaks oscillate due to the non-rephasing pathways, whereas the cross peaks (and their red-shifted ESA doublets) oscillate due to the rephasing pathways.[45] The dephasing of the LP/UP coherence arises primarily from the cavity lifetime, which is estimated to be 2.7 ps based on 92% mirror reflectivity (see SI).

Besides the coherent oscillation, we also observe waiting time amplitude changes that reflect energy transfer between the polaritons, as well as ultimate relaxation back to the ground state. Our spectra show that the polaritonic populations decay on time scales that are far longer than the ~3 ps cavity lifetime. The UP diagonal peak exhibits biexponential decay with a fast time constant of 1.4 ps and a slow timescale of roughly 10 ps. Previous work by Xiong et al.[12, 14, 40] and by Owrutsky et al.[10, 38] also show the same trend, where the slow component is expected to match the bare molecule lifetime, as the cavity signal has already decayed as that point. We attribute the 1.4 ps time constant to UP-to-LP population transfer because we see a rising component of the same time scale in the cross peak between UP and LP. By analyzing waiting-time dependent peak amplitudes, which largely reflect the population dynamics (**Fig. 4**), we can conclude qualitatively that for the regions where there is less overlap between the polariton and background spectrum (regions A and B of Fig. 4C), the residual polariton dynamics is not influenced by the background signal. For the regions of greater overlap, subtracting the background signal results in the LP diagonal (peak C in Fig. 4C) showing a reduction in the longer time decay constant while the LPUP cross peak's rise time constant becomes slower (D, Figure 4(C)). All fitting parameters for the curves shown in Fig. 4C are given in **Table S1**. Finally, the polaritonic state energies can be derived directly from the 2D spectrum at early waiting time, yielding an energy ladder diagram shown in **Fig. 4B**. In the case of ultrastrong coupling, which we do not have in the present case, there are reports of evidence of transitions between the dark and polariton manifolds.[46] For example a dark-to-UP overtone ($D_1 \rightarrow UP_2$ in our notation), would occur with an excitation frequency of 1977 cm$^{-1}$ and a detection frequency of 2011 cm$^{-1}$. Though there may be amplitude at this spectral location, it is not significant compared with the polaritonic response. We have indicated the energy differences for any dark and polaritonic transitions in Fig. 4B.

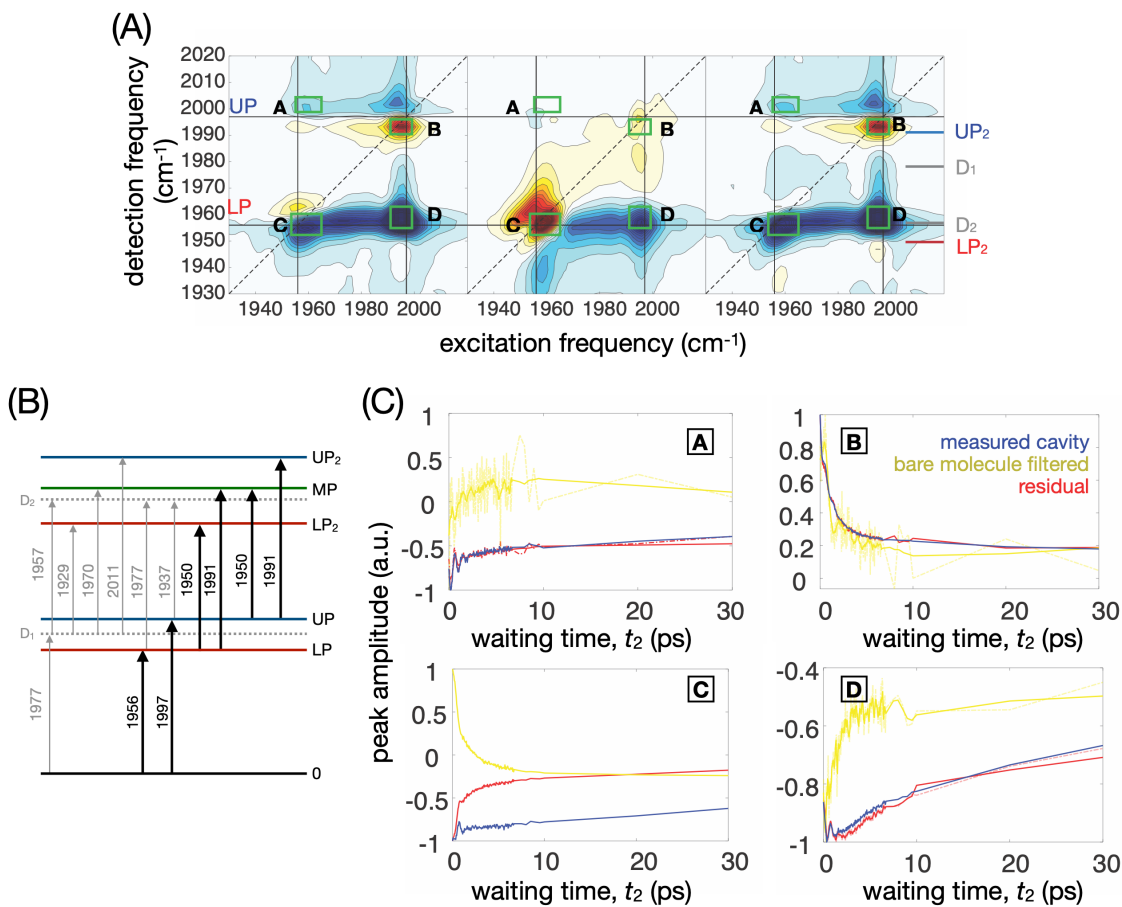

**Figure 4. (A)** Measured full cavity response (left), filtered bare molecule (middle), and subtracted residual (cavity – filtered bare molecule) at a waiting time of 0.8 ps. **(B)** Polaritonic ladder deduced from the 30 ps residual spectrum (shown in Fig. 3D, right). Gray arrows indicate transitions involving the uncoupled and dark states ($D_1$, $D_2$); bold arrows depict transitions between polariton states (LP, UP, $LP_2$, MP, and $UP_2$). **(C)** Waiting time dependence of the two diagonal (peaks C and B) and two main cross peaks (peaks A and D) as indicated in Fig. 4A. Peak A shows essentially negligible contributions from the uncoupled background, whereas the LP diagonal, peak E, shows a pronounced background contribution due to the uncoupled background. For peaks A, B and D, data are shown as moving averages (dark lines) and raw measured data (light lines); window sizes are 10 for peaks A and D, and 3 for peak F.

Regarding the role of dark states in the relaxation of polaritons, there should be two experimental signals associated with polariton-to-dark energy transfer. One feature would be a $v = 1$ to $v = 2$ ESA associated with excitation of either of the polaritons, which happens to coincide with the LP transmission. The other feature would be a stimulated emission from $v = 1$ to $v = 0$ in the uncoupled reservoir. Because that signal emits in the transmission minimum of the polariton spectrum, it will be more difficult to observe. In the subtracted spectrum, we observe a peak at (1956 cm$^{-1}$, 1977 cm$^{-1}$), corresponding to LP excitation followed by detection at the bare molecule fundamental. The sign of this LP/bare cross peak corresponds to an induced absorption (blue in our color scheme) instead of the expected stimulated emission feature (red in our color scheme). It is difficult to conclude whether this feature simply arises from the filtered bare molecule response, where it is rather prominent (Fig. 3B). The LP/bare ESA peak shape is elongated along the diagonal, which is not expected for a purely polaritonic response as the polariton features are not inhomogeneously broadened. If this feature is due to polariton-to-reservoir energy transfer, it is possible that the inhomogeneous broadening of the reservoir states would manifest as a frequency correlation. For the inhomogeneity to persist, the energy transfer to the reservoir must be faster than the spectral diffusion within the reservoir. Nevertheless, it is difficult to rationalize why the excitation frequency of the homogeneously broadened polariton would be correlated with the inhomogeneously broadened reservoir. From the bare-molecule 2D-IR we determine the spectral diffusion to occur on a 2.4 ps timescale, setting an upper bound on the energy transfer time scale. Unfortunately, it is not possible to conclude definitively that the diagonally elongated peak shape is due to energy transfer without observing the concomitant stimulated emission signal.

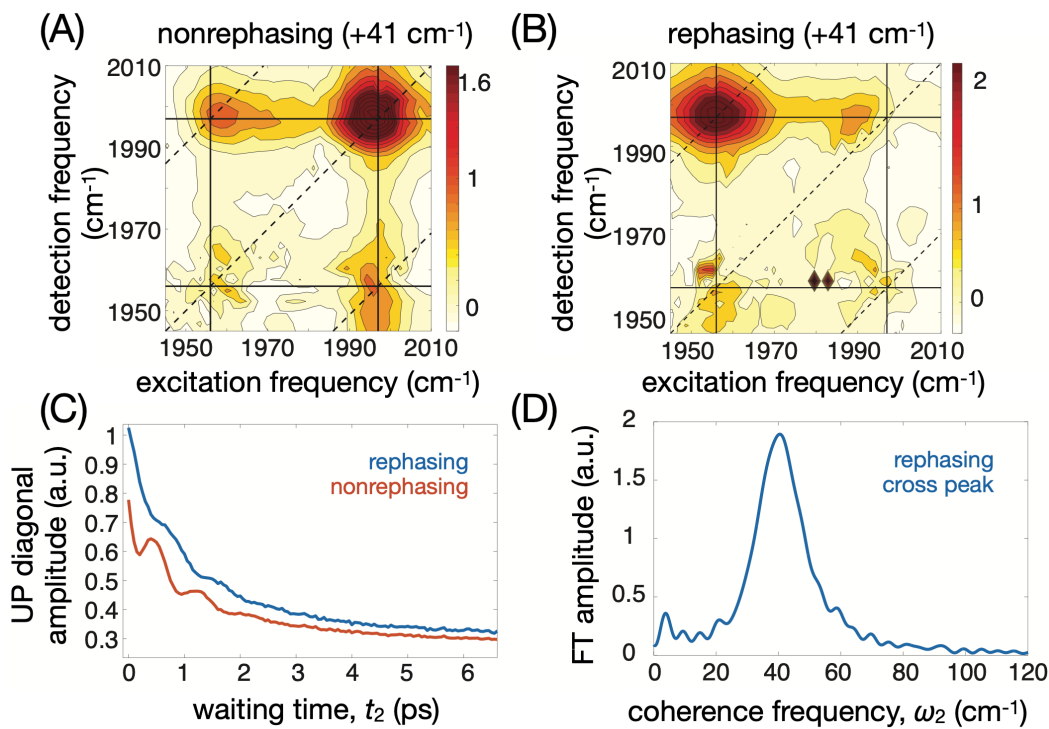

**Figure 5. (A)** Coherence map for the nonrephasing response shown at a frequency cut of +41 cm$^{-1}$. Because we analyze the complex rephasing spectrum, the coherence for the UP diagonal appears at positive frequency, whereas the LP diagonal appears at −41 cm$^{-1}$ (negative frequency coherence maps are shown in the SI). **(B)** The rephasing coherence map at +41 cm$^{-1}$, shows the upper left cross peak oscillates (the lower right cross peak oscillates with a frequency of −41 cm$^{-1}$). **(C)** Full waiting time dependence of the UP diagonal peak for the rephasing (blue) and nonrephasing (red) responses. **(D)** Absolute square Fourier transform amplitude of the rephasing response at the LP/UP cross peak indicates a single peak at 41 cm$^{-1}$, with a width of roughly 15 cm$^{-1}$, corresponding to an exponential dephasing of ~3 ps (assuming a Lorentzian spectrum).

As mentioned above, and by others,[10, 12, 38] the coherence in the waiting time is a signature of a polariton system. As is expected for a single vibrational mode, the filtered bare molecule response does not exhibit any quantum beats. Fourier transforming the cavity response with respect to $t_2$ yields a coherence frequency, $\omega_2$. Pulse shaping can be used either to isolate or suppress these coherent oscillations,[17, 38] and we have chosen to analyze them using coherence maps. Employing a 6 phase cycling (SI) scheme[47] enables direct measurement of both the real and imaginary parts of the rephasing and nonrephasing spectra. Due to the specific signs of the frequency of each pathway, this separation enables a greatly simplified coherence map that is readily assigned for each pathway. We show here (**Fig. 5**) the positive vacuum Rabi splitting (+41 cm$^{-1}$), with the negative frequency maps shown in the SI. As expected, the coherence appears on the upper diagonal for the nonrephasing and on the upper cross peak for the rephasing responses. In the rephasing LP excited/UP detected cross peak (Fig. 5D), we find a single coherence frequency at the 41 cm$^{-1}$ vacuum Rabi splitting, with a width of roughly 15 cm$^{-1}$, corresponding to an exponential dephasing of about 3 ps (assuming a Lorentzian spectrum), consistent with the ~3 ps calculated cavity lifetime (see SI).

We present a method of removing the polariton-filtered background 2D-IR spectrum that comprises roughly half of the signal amplitude in many cavity-based experiments. The dark state spectral amplitude arises from molecules located, for example, in low-field regions of the cavity, producing appreciable oscillator strength of the reservoir states. These features are inherent in vibrational polaritons, and care must be taken in interpreting polariton nonlinear spectroscopy without correcting for this potentially significant background contribution. Once corrected, the polariton 2D-IR response can be interpreted using the standard concepts of coherence, energy transfer and energy relaxation that have been developed in the multidimensional spectroscopy community.[43, 48] Compared with ordinary molecules, polaritons are immune to inhomogeneous broadening, thus enabling a subtraction procedure that eliminates the spectral diffusion contribution to the FFCF in the cavity response. While the approach we have demonstrated here can reduce the uncoupled background nonlinear response, for cases with more complex bare-molecule spectra, the method may not be sufficient. We propose that the most straightforward solution is either to (1) rely on the ultrastrong coupling where the polaritons are shifted very far from the bare molecule transitions, or (2) to ensure the active molecules are located in high-field regions of the optical cavity, thus reducing the contribution of the uncoupled molecules. Other dynamical signatures of non-polaritonic contributions may aid in identifying processes such as energy transfer to and from the dark reservoir. These findings also suggest that nanophotonic resonances may offer a notable advantage due to the potential to achieve strong coupling with single molecules.[49-50]

## ASSOCIATED CONTENT
**Supporting Information**
The Supporting Information is available free of charge at XXX.

Illustration of spatial distribution of polariton eigenmodes; Experimental methods; Effect of the spectral diffusion on long-time estimate of the uncoupled 2D-IR background; Detailed background subtraction process; Coherence maps for $\omega_2 = -41$ cm$^{-1}$; Estimate of the cavity lifetime; Fitting results for several peaks in the 2D spectrum; Calculation of molecule numbers in the cavity; and supplemental references.

## AUTHOR INFORMATION
**Corresponding Author**
   **Kevin J. Kubarych** – Department of Chemistry, University of Michigan, 930 N. University Ave, Ann Arbor, MI 48109, USA; orcid.org/0000-0003-1152-4734; Email: kubarych@umich.edu

**Authors**
   **Rong Duan** – Department of Chemistry, University of Michigan, 930 N. University Ave, Ann Arbor, MI 48109, USA; orcid.org/000-0003-1077-8134; Email: drong@umich.edu

   **Joseph N. Mastron** – Department of Chemistry, University of Michigan, 930 N. University Ave, Ann Arbor, MI 48109, USA; Department of Physics, University of Michigan, 450 Church St., Ann Arbor, MI 48109, USA; orcid.org/0000-0003-0681-2627; Email: mastron@umich.edu

**Yin Song** – Department of Physics, University of Michigan, 450 Church St., Ann Arbor, MI 48109, USA; orcid.org/0000-0001-8978-2958; Email: songyi@umich.edu

**Notes**
The authors declare no competing financial interest.

**Acknowledgements**
This work was supported by the National Science Foundation (CHE-1836529 and CHE-1955026).

**REFERENCES**
1. Deng, H.; Weihs, G.; Santori, C.; Bloch, J.; Yamamoto, Y., Condensation of semiconductor microcavity exciton polaritons. *Science* **2002,** *298*, 199-202.
2. Kena-Cohen, S.; Forrest, S. R., Room-temperature polariton lasing in an organic single-crystal microcavity. *Nat. Photon.* **2010,** *4*, 371-375.
3. Sun, Y. B.; Yoon, Y.; Steger, M.; Liu, G. Q.; Pfeiffer, L. N.; West, K.; Snoke, D. W.; Nelson, K. A., Direct measurement of polariton-polariton interaction strength. *Nat. Phys.* **2017,** *13*, 870-875.
4. Hutchison, J. A.; Schwartz, T.; Genet, C.; Devaux, E.; Ebbesen, T. W., Modifying Chemical Landscapes by Coupling to Vacuum Fields. *Angew. Chem.-Int. Edit.* **2012,** *51*, 1592-1596.
5. Thomas, A.; George, J.; Shalabney, A.; Dryzhakov, M.; Varma, S. J.; Moran, J.; Chervy, T.; Zhong, X. L.; Devaux, E.; Genet, C.; Hutchison, J. A.; Ebbesen, T. W., Ground-State Chemical Reactivity under Vibrational Coupling to the Vacuum Electromagnetic Field. *Angew. Chem.-Int. Edit.* **2016,** *55*, 11462-11466.
6. Thomas, A.; Lethuillier-Karl, L.; Nagarajan, K.; Vergauwe, R. M. A.; George, J.; Chervy, T.; Shalabney, A.; Devaux, E.; Genet, C.; Moran, J.; Ebbesen, T. W., Tilting a ground-state reactivity landscape by vibrational strong coupling. *Science* **2019,** *363*, 615-619.
7. Pang, Y. T.; Thomas, A.; Nagarajan, K.; Vergauwe, R. M. A.; Joseph, K.; Patrahau, B.; Wang, K. D.; Genet, C.; Ebbesen, T. W., On the Role of Symmetry in Vibrational Strong Coupling: The Case of Charge-Transfer Complexation. *Angew. Chem.-Int. Edit.* **2020,** *59*, 10436-10440.
8. Vergauwe, R. M. A.; Thomas, A.; Nagarajan, K.; Shalabney, A.; George, J.; Chervy, T.; Seidel, M.; Devaux, E.; Torbeev, V.; Ebbesen, T. W., Modification of Enzyme Activity by Vibrational Strong Coupling of Water. *Angew. Chem.-Int. Edit.* **2019,** *58*, 15324-15328.
9. Lather, J.; George, J., Improving Enzyme Catalytic Efficiency by Co-operative Vibrational Strong Coupling of Water. *J. Phys. Chem. Lett.* **2021,** *12*, 379-384.
10. Dunkelberger, A. D.; Spann, B. T.; Fears, K. P.; Simpkins, B. S.; Owrutsky, J. C., Modified relaxation dynamics and coherent energy exchange in coupled vibration-cavity polaritons. *Nat. Commun.* **2016,** *7*, 13504.
11. Dunkelberger, A. D.; Davidson, R. B.; Ahn, W.; Simpkins, B. S.; Owrutsky, J. C., Ultrafast Transmission Modulation and Recovery via Vibrational Strong Coupling. *J. Phys. Chem. A* **2018,** *122*, 965-971.
12. Xiang, B.; Ribeiro, R. F.; Dunkelberger, A. D.; Wang, J. X.; Li, Y. M.; Simpkins, B. S.; Owrutsky, J. C.; Yuen-Zhou, J.; Xiong, W., Two-dimensional infrared spectroscopy of vibrational polaritons. *Proc. Natl. Acad. Sci. U. S. A.* **2018,** *115*, 4845-4850.

13. Dunkelberger, A. D.; Grafton, A. B.; Vurgaftman, I.; Soykal, O. O.; Reinecke, T. L.; Davidson, R. B.; Simpkins, B. S.; Owrutsky, J. C., Saturable Absorption in Solution-Phase and Cavity-Coupled Tungsten Hexacarbonyl. *ACS Photonics* **2019,** *6*, 2719-2725.
14. Xiang, B.; Ribeiro, R. F.; Chen, L. Y.; Wang, J. X.; Du, M.; Yuen, O. E. Z.; Xiong, W., State-Selective Polariton to Dark State Relaxation Dynamics. *J. Phys. Chem. A* **2019,** *123*, 5918-5927.
15. Xiang, B.; Ribeiro, R. F.; Li, Y. M.; Dunkelberger, A. D.; Simpkins, B. B.; Yuen-Zhou, J.; Xiong, W., Manipulating optical nonlinearities of molecular polaritons by delocalization. *Science Advances* **2019,** *5*, eaax5196.
16. Xiang, B.; Ribeiro, R. F.; Du, M.; Chen, L. Y.; Yang, Z. M.; Wang, J. X.; Yuen-Zhou, J.; Xiong, W., Intermolecular vibrational energy transfer enabled by microcavity strong light-matter coupling. *Science* **2020,** *368*, 665-667.
17. Yang, Z. M.; Xiang, B.; Xiong, W., Controlling Quantum Pathways in Molecular Vibrational Polaritons. *ACS Photonics* **2020,** *7*, 919-924.
18. Xiang, B.; Wang, J. X.; Yang, Z. M.; Xiong, W., Nonlinear infrared polaritonic interaction between cavities mediated by molecular vibrations at ultrafast time scale. *Science Advances* **2021,** *7*, eabf6397.
19. Xiang, B.; Xiong, W., Molecular vibrational polariton: Its dynamics and potentials in novel chemistry and quantum technology. *J. Chem. Phys.* **2021,** *155*, 050901.
20. Shalabney, A.; George, J.; Hutchison, J.; Pupillo, G.; Genet, C.; Ebbesen, T. W., Coherent coupling of molecular resonators with a microcavity mode. *Nat. Commun.* **2015,** *6*, 5981.
21. Martinez-Martinez, L. A.; Ribeiro, R. F.; Campos-Gonzalez-Angulo, J.; Yuen-Zhou, J., Can Ultrastrong Coupling Change Ground-State Chemical Reactions? *ACS Photonics* **2018,** *5*, 167-176.
22. Ribeiro, R. F.; Martinez-Martinez, L. A.; Du, M.; Campos-Gonzalez-Angulo, J.; Yuen-Zhou, J., Polariton chemistry: controlling molecular dynamics with optical cavities. *Chem. Sci.* **2018,** *9*, 6325-6339.
23. Campos-Gonzalez-Angulo, J. A.; Ribeiro, R. F.; Yuen-Zhou, J., Resonant catalysis of thermally activated chemical reactions with vibrational polaritons. *Nat. Commun.* **2019,** *10*, 4685.
24. Du, M.; Ribeiro, R. F.; Yuen-Zhou, J., Remote Control of Chemistry in Optical Cavities. *Chem* **2019,** *5*, 1167-1181.
25. Yuen-Zhou, J.; Menon, V. M., Polariton chemistry: Thinking inside the (photon) box. *Proc. Natl. Acad. Sci. U. S. A.* **2019,** *116*, 5214-5216.
26. Wasielewski, M. R.; Forbes, M. D. E.; Frank, N. L.; Kowalski, K.; Scholes, G. D.; Yuen-Zhou, J.; Baldo, M. A.; Freedman, D. E.; Goldsmith, R. H.; Goodson, T.; Kirk, M. L.; McCusker, J. K.; Ogilvie, J. P.; Shultz, D. A.; Stoll, S.; Whaley, K. B., Exploiting chemistry and molecular systems for quantum information science. *Nat. Rev. Chem.* **2020,** *4*, 490-504.
27. Li, T. E.; Nitzan, A.; Subotnik, J. E., On the origin of ground-state vacuum-field catalysis: Equilibrium consideration. *J. Chem. Phys.* **2020,** *152*, 234107.
28. Li, T. E.; Subotnik, J. E.; Nitzan, A., Cavity molecular dynamics simulations of liquid water under vibrational ultrastrong coupling. *Proc. Natl. Acad. Sci. U. S. A.* **2020,** *117*, 18324-18331.
29. Li, T. E.; Nitzan, A.; Subotnik, J. E., Cavity molecular dynamics simulations of vibrational polariton-enhanced molecular nonlinear absorption. *J. Chem. Phys.* **2021,** *154*, 094124.
30. Li, X. Y.; Mandal, A.; Huo, P. F., Cavity frequency-dependent theory for vibrational polariton chemistry. *Nat. Commun.* **2021,** *12*, 1315.

31. Sidler, D.; Ruggenthaler, M.; Appel, H.; Rubio, A., Chemistry in Quantum Cavities: Exact Results, the Impact of Thermal Velocities, and Modified Dissociation. *J. Phys. Chem. Lett.* **2020,** *11*, 7525-7530.
32. Avramenko, A. G.; Rury, A. S., Quantum Control of Ultrafast internal Conversion Using Nanoconfined Virtual Photons. *J. Phys. Chem. Lett.* **2020,** *11*, 1013-1021.
33. Jaynes, E. T.; Cummings, F. W., Comparison of Quantum and Semiclassical Radiation Theories with Application to Beam Maser. *Proceedings of the Ieee* **1963,** *51*, 89-109.
34. Tavis, M.; Cummings, F. W., Approximate Solutions for an N-Molecule-Radiation-Field Hamiltonian. *Phys. Rev.* **1969,** *188*, 692-695.
35. Vurgaftman, I.; Simpkins, B. S.; Dunkelberger, A. D.; Owrutsky, J. C., Negligible Effect of Vibrational Polaritons on Chemical Reaction Rates via the Density of States Pathway. *J. Phys. Chem. Lett.* **2020,** *11*, 3557-3562.
36. Herrera, F.; Spano, F. C., Absorption and photoluminescence in organic cavity QED. *Phys. Rev. A* **2017,** *95*, 053867.
37. Crum, V. F.; Casey, S. R.; Sparks, J. R., Photon-mediated hybridization of molecular vibrational states. *Phys. Chem. Chem. Phys.* **2018,** *20*, 850-857.
38. Grafton, A. B.; Dunkelberger, A. D.; Simpkins, B. S.; Triana, J. F.; Hernandez, F. J.; Herrera, F.; Owrutsky, J. C., Excited-state vibration-polariton transitions and dynamics in nitroprusside. *Nat. Commun.* **2021,** *12*, 214.
39. George, J.; Chervy, T.; Shalabney, A.; Devaux, E.; Hiura, H.; Genet, C.; Ebbesen, T. W., Multiple Rabi Splittings under Ultrastrong Vibrational Coupling. *Phys. Rev. Lett.* **2016,** *117*, 153601.
40. Ribeiro, R. F.; Campos-Gonzalez-Angulo, J. A.; Giebink, N. C.; Xiong, W.; Yuen-Zhou, J., Enhanced optical nonlinearities under collective strong light-matter coupling. *Phys. Rev. A* **2021,** *103*, 063111.
41. Brookes, J. F.; Slenkamp, K. M.; Lynch, M. S.; Khalil, M., Effect of Solvent Polarity on the Vibrational Dephasing Dynamics of the Nitrosyl Stretch in an Fe-II Complex Revealed by 2D IR Spectroscopy. *J. Phys. Chem. A* **2013,** *117*, 6234-6243.
42. Hernandez, F.; Herrera, F., Multi-level quantum Rabi model for anharmonic vibrational polaritons. *J. Chem. Phys.* **2019,** *151*, 144116.
43. Hamm, P.; Zanni, M. T., *Concepts and Methods of 2D Infrared Spectroscopy*. Cambridge University Press: New York, 2011.
44. Duan, R.; Mastron, J. N.; Song, Y.; Kubarych, K. J., Direct comparison of amplitude and geometric measures of spectral inhomogeneity using phase-cycled 2D-IR spectroscopy. *J. Chem. Phys.* **2021,** *154*, 174202.
45. Nee, M. J.; Baiz, C. R.; Anna, J. M.; McCanne, R.; Kubarych, K. J., Multilevel vibrational coherence transfer and wavepacket dynamics probed with multidimensional IR spectroscopy. *J. Chem. Phys.* **2008,** *129*, 084503.
46. DelPo, C. A.; Kudisch, B.; Park, K. H.; Khan, S. U. Z.; Fassioli, F.; Fausti, D.; Rand, B. P.; Scholes, G. D., Polariton Transitions in Femtosecond Transient Absorption Studies of Ultrastrong Light-Molecule Coupling. *J. Phys. Chem. Lett.* **2020,** *11*, 2667-2674.
47. Yan, S. X.; Tan, H. S., Phase cycling schemes for two-dimensional optical spectroscopy with a pump-probe beam geometry. *Chem. Phys.* **2009,** *360*, 110-115.
48. Kiefer, L. M.; Kubarych, K. J., Two-dimensional infrared spectroscopy of coordination complexes: From solvent dynamics to photocatalysis. *Coord. Chem. Rev.* **2018,** *372*, 153-178.

49. Chikkaraddy, R.; de Nijs, B.; Benz, F.; Barrow, S. J.; Scherman, O. A.; Rosta, E.; Demetriadou, A.; Fox, P.; Hess, O.; Baumberg, J. J., Single-molecule strong coupling at room temperature in plasmonic nanocavities. *Nature* **2016,** *535*, 127-130.
50. Cohn, B.; Das, K.; Basu, A.; Chuntonov, L., Infrared Open Cavities for Strong Vibrational Coupling. *J. Phys. Chem. Lett.* **2021,** *12*, 7060-7066.

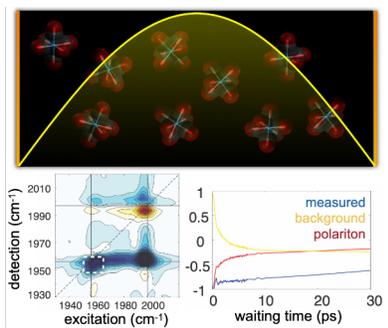

TOC Graphic

**Supporting Information**

# Isolating Polaritonic 2D-IR Transmission Spectra

Rong Duan[(1)], Joseph N. Mastron[(1,2)], Yin Song[(2,#)], Kevin J. Kubarych[(1)]*
[1] Department of Chemistry, University of Michigan, 930 N. University Ave, Ann Arbor, MI 48109
[2] Department of Physics, University of Michigan, 430 Church Ave, Ann Arbor, MI 48109
[#] Present address: Beijing Institute of Technology, 5 South Zhong Guan Cun St., Beijing, China
*E-mail: kubarych@umich.edu

**Contents**

**S1.** Illustration of spatial distribution of polariton eigenmodes
**S2.** Experimental method: 2D-IR spectrometer
**S3.** Effect of spectral diffusion on long-time estimate of the uncoupled 2D-IR background
**S4.** Detailed background subtraction process
**S5.** Coherence maps for $\omega_2$ = –41 cm$^{-1}$
**S6.** Estimate of the cavity lifetime
**S7.** Fitting results for several peaks in the 2D spectrum
**S8.** Calculation of molecule numbers in the cavity
**S9.** References

## S1. Illustration of spatial distribution of polariton eigenmodes

Cavity mode resonances arise due to the boundary conditions imposed by the mirrors. Because the coupling to the cavity is proportional to the field amplitude, molecules near the mirrors (or any nodes in higher-index optical modes) will experience reduced coupling. These molecules can be included in a Tavis-Cummings Hamiltonian[1] by simply making the coupling to the cavity depend on each molecule's longitudinal position in the cavity. The result is that the polariton eigenvectors have coefficients that follow the field (**Fig. S1**). The locations of the dark states also follow the field, but the uncoupled molecules are located near the mirror interfaces. The uncoupled contribution, as well as the dark states, are responsible for the substantial non-polaritonic background.

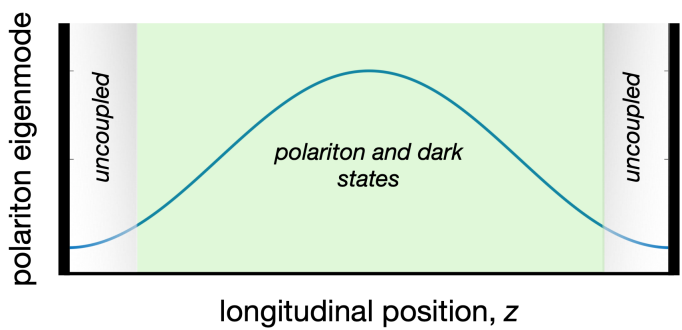

**Figure S1.** Eigenvector distribution of a polariton in a cavity. The blue line shows the squared modulus of the molecular part of the polariton eigenvector (i.e. $|c_j|^2$). Near the mirrors, there is little participation from molecules in both the polariton and dark state eigenvectors. In the regions of higher cavity mode field, there are dark states and the polariton states.

## S2. Experimental method: 2D-IR spectrometer and cavity fabrication

The 2D-IR experiment is performed using a pump probe geometry.[2-3] The Spectra-Physics Solstice Ace amplifier generates 100-fs, 4.2-mJ, 800-nm pulses at a 1-kHz repetition rate. This pulse train pumps an optical parametric amplifier (OPA, Light Conversion TOPAS-Twins). The OPA output is sent into a home-built collinear difference frequency generation stage using a $AgGaS_2$ crystal to generate 9-µJ mid-IR pulses centered at 5050 nm. Using a $CaF_2$ wedge, we direct 95% of the beam into a germanium acoustic-optic modulator-based pulse shaper (PhaseTech Spectroscopy) to produce the phase-controlled excitation pulse pair. The remaining 5% serves as the probe pulse. Both pump and probe pulses are ~100 fs. A 15 cm effective focal length parabolic mirror focuses the two beams at the sample position, and only the probe is detected using a 0.3-m spectrograph (Horiba-Yvon, iHR320) equipped with a 64-pixel HgCdTe (MCT) IR detector array.

    2D spectra were recorded at a fixed waiting time ($t_2$) delay while scanning $t_1$ from 0 to 6 ps with 10 fs time steps and a rotation frame set at 1700 cm$^{-1}$. A 6 phase cycling scheme is used to separate the rephasing and nonrephasing spectra[4] while also allowing for subtraction of the scattered light. Raw data is windowed using a Hann function and zero padded to a computed spectral resolution of 0.8 cm$^{-1}$. 2D-IR spectra were collected at waiting time ($t_2$) points from -2 ps to 0 ps in steps of 500 fs, from 0 to 7 ps in steps of 50 fs, from 7 to 10 ps in steps of 500 fs, and from 10 to 30 ps in steps of 10 ps.

    Cavity mirrors (coated by Universal Thin Film Lab) are coated with magnesium fluoride ($MgF_2$) on 25-mm diameter, 3-mm thick calcium fluoride ($CaF_2$) windows. The reflectivity at the region of interest is

~92%. The coarse cavity length was set using a PTFE spacer of known thickness, and then finely adjusted to shift the Fabry-Pérot transmission bands using a spectrometer. In terms of sample preparation for the experiments, the same concentration of W(CO)$_6$ (40 mM) was filled inside the 25 μm cavity for the polariton experiments and a 6 μm spacer was used for the bare molecule experiments.

## S3. Effect of spectral diffusion on long-time estimate of the uncoupled 2D-IR background

Due to inhomogeneous broadening and spectral diffusion, 2D-IR spectra of molecules in polar solvents typically exhibit dynamical line shapes that change as a function of the waiting time ($t_2$). We show here that this dynamical line shape causes the majority of the spectral features attributed to the polariton response in the work by Grafton *et al.* on sodium nitroprusside.[5] To simulate their results, we compute two 2D-IR absorptive spectra (using standard response functions),[6] one at an early waiting time (i.e. before significant spectral diffusion occurs) and another at a later waiting time when the response appears homogeneous (i.e. spectral diffusion is complete). We then filter these bare-molecule 2D spectra by the reported polariton transmission (data shown in **Fig. S2A** are taken from Ref. S5). Both 2D-IR spectra are multiplied by the filter function (**Fig. S2B**) to produce two filtered bare-molecule responses (**Fig. S2B,D**). We subtract the late time response from the early time response, to give the residual spectra (**Fig. S2E,F**), which resemble the polariton spectrum reported in Ref. S5.

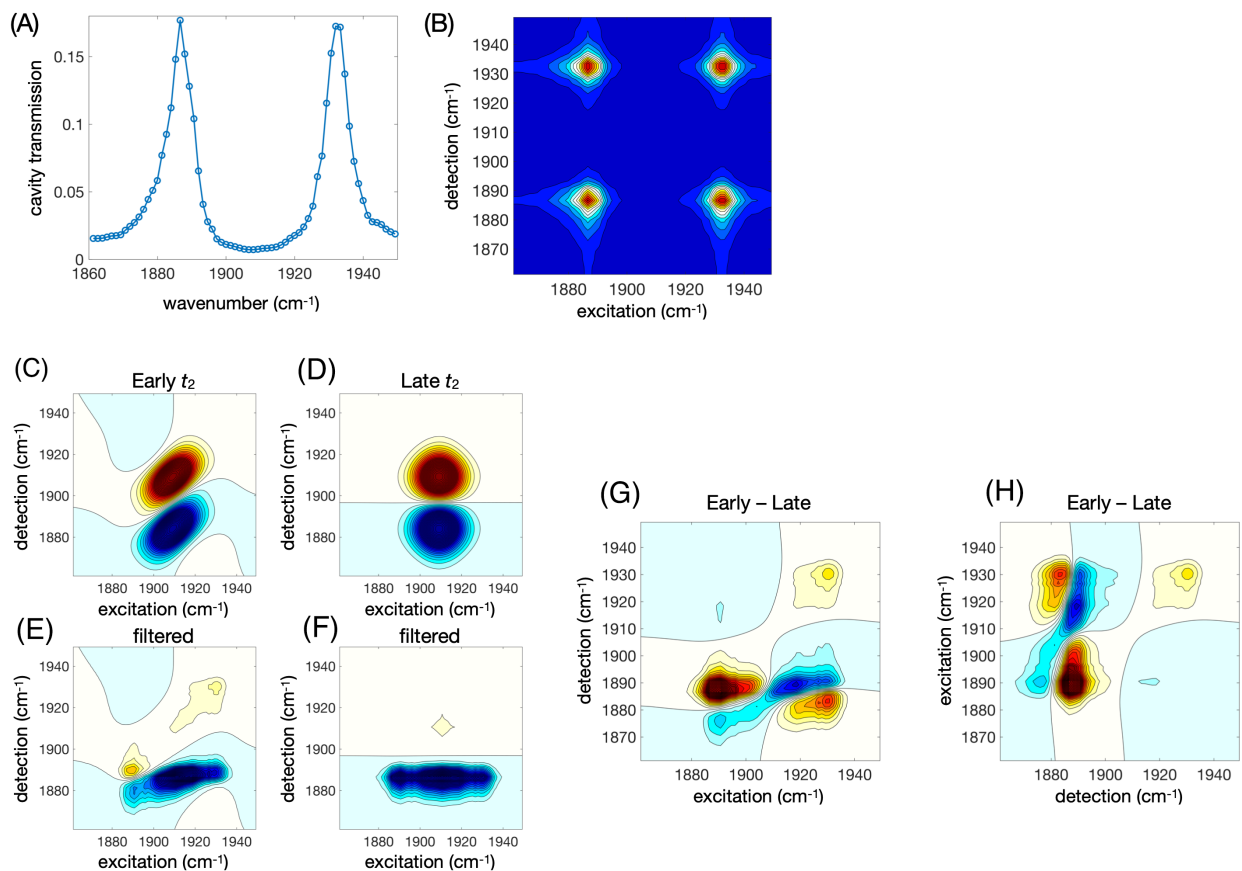

**Figure S2. (A)** Transmission spectrum (data extracted from Ref. S5) used to filter the simulated 2D-IR response. **(B)** 2D transmission filter. **(C)** Simulated early waiting time ($t_2$) absorptive 2D-IR spectrum resembling the results of Khalil et al. for sodium nitroprusside. **(D)** Simulated late waiting time 2D-IR spectrum. **(E)** Filtered early time 2D-IR spectrum. **(F)** Filtered late-time 2D-IR spectrum. **(G)** Late time, filtered 2D-IR spectrum subtracted from the early-time filtered 2D-IR spectrum. **(H)** Same difference spectrum plotted using the convention where the excitation and detection axes are swapped (for easier comparison with Ref. S5).

**S4. Detailed background subtraction process**

We have devised an approach for removing the background uncoupled 2D-IR response by using the spectral diffusion as a feedback signal. Because we have used a 6 phase cycling scheme, we can construct the complex rephasing and nonrephasing signals. We choose the weighting factor with which we scale the bare molecule 2D-IR signal such that we eliminate the spectral diffusion signal in the residual polariton spectrum (most pronounced in the LPLP diagonal peak, see Fig. 2D of the main text). Due to the coherent oscillations that appear in the FFCF (expected for polaritons), we use the signal after 2 ps to set the weight parameter. Fig. S3 shows that with an appropriate choice of the weight, we can remove the spectral diffusion and leave only the coherent oscillation caused by impulsive excitation of the two polariton bands. We show how the square of the residual (averaged from $t_2$ = 2 ps to $t_2$ = 6.7 ps) depends on the weight parameter. We choose the value that minimizes this squared difference from the expected value of zero. Due to slight differences in alignment between the cavity and non-cavity measurements, we find that we need to shift the $t_2$ of the cavity 300 fs to earlier time relative to the non-cavity measurements. The optimization process is summarized in Fig. S3.

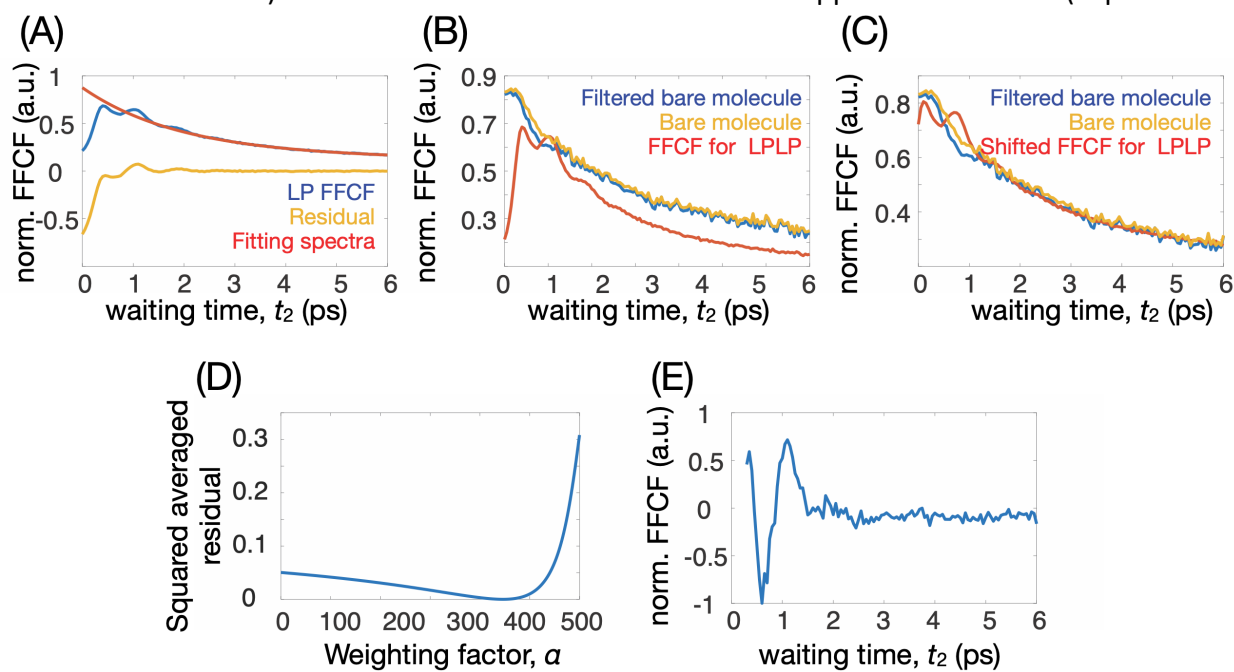

**Figure S3.** Using the FFCF of the LP to find the weighting parameter. (A) Waiting time dependence of the idealized FFCF of the LP band (yellow) obtained by subtracting the fitted molecule background (red) from normalized cavity LP normalized FFCF (blue). (B) $t_2$ = 0 comparison using the waiting time dependent FFCF from the filtered bare molecule (blue), bare molecule (yellow), and cavity LP (yellow). (C) Shifting the cavity FFCF forward by 300 fs and adding an offset of 0.12 to show that the LP FFCF is correctly timed. (D) Varying the weighting parameter gives a squared, average LP residual after 2 ps that reaches a minimum near 350 (arbitrary units). (E) The LP FFCF obtained by subtracting the weighted bare-molecule spectrum at each waiting time shows the elimination of spectral diffusion.

## S5. Coherence maps for $\omega_2 = -41$ cm$^{-1}$

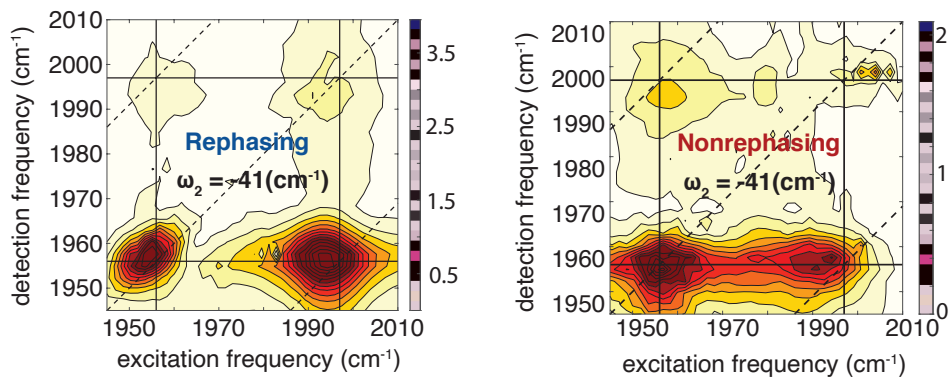

**Figure S4.** Coherence maps (absolute value amplitude) for the rephasing and nonrephasing signals at a coherence frequency of -41 cm$^{-1}$.

Using a 6 phase cycling scheme to collect the data, we can obtain both real and imaginary parts of the rephasing and nonrephasing spectra. As a result of Fourier transformation with respect to $t_2$, rephasing cross peaks with excitation frequency greater than the detection frequency will appear with negative coherence frequency. Likewise, the coherence located at the lower nonrephasing diagonal peak also appears with a negative frequency. Absolute value contour plots (**Fig. S4**) show the amplitude of the coherence maps.

## S6. Estimate of the cavity lifetime
For a Fabry-Pérot resonator with loss arising from the mirror reflectivity, the cavity lifetime is given by:[7]

$$\tau_{cav} = T/(1-R)$$

Where T is the roundtrip time and R is the reflectivity. For a 25 µm cavity, the round-trip time is 50 µm/((0.3 µm/fs)/1.3) = 216 fs (using an index of refraction of 1.3). Therefore the estimated cavity lifetime is 216/(1-0.92) = 2.7 ps.

## S7. Fitting results for several peaks in the 2D spectrum
We isolate the polariton spectrum by subtracting the filtered bare molecule 2D-IR response from the as-measured cavity response:

$$S_{Residual} = S_{cavity} - \alpha S_{filtered\ bare\ molecule}$$

S5

|  | Measured cavity | | Bare molecule filtered | | residual | |
|---|---|---|---|---|---|---|
| peak | $\tau_1$(ps) | $\tau_2$(ps) | $\tau_1$(ps) | $\tau_2$(ps) | $\tau_1$(ps) | $\tau_2$(ps) |
| A | 2.32±1.28 | 97.73±120.97 | - | - | 7.67±3.87 | - |
| B | 0.31±0.05 | 1.6±0.20 | 0.58±0.20 | 8.1±16.40 | 1.30±0.10 | 46.35±204.55 |
| C | 0.17±0.01 | 102.7±283.70 | 0.60±0.07 | 3.58±0.95 | 0.21±0.04 | 3.5±0.48 |
| D | 1.09±0.45 | 10.5±3.37 | 1.70±0.36 | - | 1.02±0.37 | 11.56±3.15 |

**Table S1.** Table for fitting parameters. Waiting time dependent population dynamics of peaks A, D, E, and F (as labeled in Fig. S4) from the as-measured cavity, filtered bare molecule, and residual spectra.

The waiting time dependence of peaks B, C and D are fit to the equation:

$$ae^{-\frac{t_2}{\tau_1}} + a_2 e^{-\frac{t_2}{\tau_2}} + c$$

In the as-measured cavity response, peak A is fit to a biexponential without a constant offset:

$$ae^{-\frac{t_2}{\tau_1}} + a_2 e^{-\frac{t_2}{\tau_2}}$$

The filtered bare molecule spectrum at peak F can be fit to a single exponential. The filtered bare molecule at peak A has too small amplitude to yield a fit. Error bars in the plots are the 95% confidence interval, and a considerable amount of the fitting error results from the coherent modulations of the peaks that inevitably arise from impulsive nonlinear spectroscopy of coupled transitions.

### S8. Calculation of molecule numbers in the cavity
We encapsulated a 40 mM solution inside 25 mm diameter cavity mirrors with a 25 µm spacer. This will give us a volume of $1.2 \times 10^{-8}$ m$^3$, which is $1.2 \times 10^{-5}$ L. With a concentration of 40 mmol/L, we have a total of $5 \times 10^{-7}$ moles, or $3 \times 10^{17}$ molecules.

### S9. References

1. Tavis, M.; Cummings, F. W., Approximate Solutions for an N-Molecule-Radiation-Field Hamiltonian. *Phys. Rev.* **1969,** *188*, 692-695.
2. Myers, J. A.; Lewis, K. L. M.; Tekavec, P. F.; Ogilvie, J. P., Two-color two-dimensional Fourier transform electronic spectroscopy with a pulse-shaper. *Opt. Express* **2008,** *16*, 17420-17428.
3. Shim, S. H.; Strasfeld, D. B.; Ling, Y. L.; Zanni, M. T., Automated 2D IR spectroscopy using a mid-IR pulse shaper and application of this technology to the human islet amyloid polypeptide. *Proc. Natl. Acad. Sci. U. S. A.* **2007,** *104*, 14197-14202.
4. Yan, S. X.; Tan, H. S., Phase cycling schemes for two-dimensional optical spectroscopy with a pump-probe beam geometry. *Chem. Phys.* **2009,** *360*, 110-115.

5. Grafton, A. B.; Dunkelberger, A. D.; Simpkins, B. S.; Triana, J. F.; Hernandez, F. J.; Herrera, F.; Owrutsky, J. C., Excited-state vibration-polariton transitions and dynamics in nitroprusside. *Nat. Commun.* **2021,** *12*.
6. Hamm, P.; Zanni, M. T., *Concepts and Methods of 2D Infrared Spectroscopy*. Cambridge University Press: New York, 2011.
7. Renk, K. F., Fabry–Perot Resonator. In *Basics of Laser Physics: For Students of Science and Engineering*, Springer Berlin Heidelberg: Berlin, Heidelberg, 2012; pp 43-54.